\pdfoutput=1
\documentclass{aa}  
\usepackage{graphicx}
\usepackage[varg]{txfonts}
\begin{document}
   \title{High-Fidelity VLA Imaging of the Radio Structure of 3C273}


   \author{R. A. Perley
          \inst{1}
          \and
          K. Meisenheimer\inst{2}
          }

   \offprints{R. Perley}

   \institute{National Radio Astronomy Observatory\\
              Socorro NM 87801\\
              \email{rperley@nrao.edu}
         \and
             Max Planck Instit\"ut fur Astronomie\\
             Heidelberg, Germany
                          }

   \date{}

 
  \abstract
  {3C273, the nearest bright quasar, comprises a strong nuclear core
    and a bright, one-sided jet extending $\sim$23 arcseconds to the
    SW.  The source has been the subject of imaging campaigns in all
    wavebands.  Extensive observations of this source have been made
    with the Very Large Array and other telescopes as part of a campaign to
    understand the jet emission mechanisms.  Partial results from the
    VLA radio campaign have been published, but to date, the complete
    set of VLA imaging results has not been made available.  }
  {We have utilized the VLA to determine the radio structure of 3C273
    in Stokes I, Q, and U, over the widest possible frequency and
    resolution range.}
  {The VLA observed the source in all four of its configurations, and
    with all eight of its frequency bands, spanning 73.8 MHz to 43
    GHz.  The data were taken in a pseudo-spectral line mode to
    minimize the VLA's correlator errors, and were fully calibrated
    with subsequent self-calibration techniques to maximise image
    fidelity.}
  {Images in Stokes parameters I, Q, and U, spanning a resolution
    range from 6 arcseconds to 88 milliarcseconds are presented.
    Spectral index images, showing the evolution of the jet component
    are shown.  Polarimetry demonstrates the direction of the magnetic
    fields responsible for the emission, and rotation measure maps
    show the RM to be very small with no discernible trend along or
    across the jet.  This paper presents a small subset of these
    images to demonstrate the major characteristics of the source
    emission. A library of all $\sim$ 500 images has been made
    available for open, free access by interested parties.}
   {}


   \maketitle
%

\section{Introduction}

3C273 (J1229+0203), the first identified quasar \citep{S63}, is
one of the closest and most luminous of all quasars. Imaging of this
source shows that at all wavebands, 3C273 comprises a bright,
flat-spectrum nuceus with highly variable flux density, and a
one-sided, highly polarized, narrow jet extending $\sim$23 arcseconds
to the SW of the nucleus.  Since its discovery, and due to its
relative proximity ($z=0.158$, so 3.4 arcsecond $\sim$ 1 kpc) and
angular size, 3C273 has been observed by a wide range of instruments
from long-wavelength radio through X-ray.  

3C273 was the target of an extensive and comprehensive observational
campaign spanning many wavebands and utilizing many telescopes from
1995 through 2005.  Key results from this campaign are to be found in
\citet{Jester05}, and references within, which presents results from
observations made with the VLA from 8 through 43 GHz, along with HST
observations from the near-ultraviolet through near-infrared.
However, the radio images shown in this paper are a small selection of
those available from the suite of VLA observations of this source.

The purpose of this paper is to present an overview of the key
results from all the VLA imaging of 3C273, from the data taken between
1987 and 1999.  

\section{Observations}

The results presented here are from three separate observational
campaigns utilizing the Very Large Array\footnote{The National Radio
  Astronomy Observatory is a facility of the National Science
  Foundation operated under cooperative agreement by Associated
  Universities, Inc.}, each with different goals.  The first, taken in
1987, 1991 and 1999 under project codes AP134, AB608, and AB916 was
intended to search for changes in the structure of the radio jet over
the 14-year span of time.  The second, in 1995 through 1997, under
project codes AR334 and AR371, were part of a full coverage study of
the jet emission properties.  Neither of these projects included
observations from the low frequency bands, at 73.8 and 327 MHz. The
results included in this paper from these bands are taken from project
AK461, whose goal was to demonstrate the imaging capability of the VLA
at low frequencies by observing a selection of 3C sources, including
3C273.

The observing details for all these projects are given in
Table~\ref{table:1}.  To minimize the baseline-based errors (`closure
errors') which limit image fidelity for very bright sources such as
3C273, we employed the spectral line mode `1A', providing 64 channels
in a single correlation over 12.5 MHz bandiwdth, or `PA' with a
bandwidth of 12.5 MHz, providing all four complex correlations with 16
channels each.  The low frequency observations were taken in a dual
frequency mode, providing only the parallel hand correlations in
spectral line mode `4', with 3.125 MHz bandwidth at 327 MHz, and 1.56
MHz bandwidth at 73.8 MHz.
\begin{table}
\begin{minipage}[t]{\columnwidth}
\caption{VLA Observing Log}
\label{table:1}
\centering
\begin{tabular}{c c c c c}
  \hline\hline
  Project&Date&Config.&Bands\footnote{4: 73.8 MHz, P: 327 MHz, L: 1365 MHz, C: 4885 MHz, X: 8415 MHz, U:
    14965 MHz, K: 22485 MHz, Q: 43315 MHz}&Duration(hr)\\
  \hline
  AP134&07 Sep 1987&A&C, U       &12\\
  AP134&10 Sep 1987&A&C, U       &12\\
  AB608&04 Jul 1991&A&C, U       &12\\
  AB608&14 Jul 1991&A&C, U       &12\\
  AR334&09 Jul 1995&A&L,C,X,U,K,Q&12\\
  AR334&13 Oct 1995&B&L,C,X,U,K,Q&12\\
  AR334&01 Mar 1996&C&L,C,X,U,K,Q&12\\
  AR371&08 Nov 1997&D&L,C,X,U,K,Q&6\\
  AK461&07 Mar 1998&A&4,P        &24\\
  AB916&19 Jun 1999&A&C, U       &10\\
  \hline
\end{tabular}
\end{minipage}
\end{table}

Because 3C273 is a very strong radio source with a nuclear core
suitable for calibration, a special calibration regimen was adopted.
At the lower frequencies (1365 MHz, 4885 MHz, and 8415 MHz), where the
jet emission is significant and the array resolution lower, phase and
amplitude calibration were accomplished with the nearby unresolved
source J1150$-$0023.  At the higher frequency bands (14965 MHz, 22485
MHz, and 43315 MHz), this secondary calibrator was not necessary, and
the nuclear component of 3C273 itself provided the calibration.

The observations for the primary projects AR334 and AR371 were arranged
on an hourly cycle, within which the first 17 minutes were used for
the observations at the lowest three frequencies (including the
calibrator), followed by 10 minutes of 14965 MHz, 12 minutes at 22485
MHz, and 15 minutes at 43315 MHz. This variation of duration with
frequency was done to offset the loss of sensitivity at the higher
frequency bands.  Referenced pointing, a technique which determines
the local pointing offsets, was done utilizing 3C273 at X-band,
with the corrections applied to the subsequent observations at higher
frequencies.

The VLA's 40 -- 50 GHz system (Q-band) was being implemented during this
program, with the result that only 9, 11, 13, and 13 antennas were outfitted
for the A, B, C, and D configuration observations, respectively.  As this
system was in its early stages, and because we did not expect to be able to
accomplish detailed structure studies with the incomplete system, we observed
only every other hour at this band.  

The flux density scale was set through an observation of 3C286, the
VLA's primary flux density calibrator, at the end of each observing
run.  The values used were from \citet{PB13}, which are based on the
\citet{Baars} for the frequencies below 2 GHz, and
on quantitative emission models of Mars for the high frequencies.  The
values utilized are shown in Table~\ref{table:2}.

\begin{table}
\caption{Flux Density of 3C286}
\label{table:2}
\centering
\begin{tabular}{c c c}
\hline\hline
Band&Freq&Flux Density\\
&MHz&Jy\\
\hline
P&  327 &25.0\\
L& 1365 &15.02\\
C& 4885&7.410\\
X& 8415 &5.197\\
U&14915&3.455\\
K&22485&2.574\\
Q&43315&1.537\\
\hline
\end{tabular}
\end{table}

Calibration of the data was done using the {\tt AIPS} software package, using the
following procedure:
\begin{enumerate}
\item Quick inspection of the amplitude data was used to remove obviously bad
  data from `dead' antennas.  
\item The bandpass functions for the polarizations observed were determined
  using 3C273 itself, using BPASS.
\item The spectral channel data were corrected by the bandpass, and collapsed
  to a single `pseudo-continuum' channel, using SPLAT.  The result is a
  single-frequency, multi-source database, similar to what the continuum
  system provides, but with much reduced baseline-dependent (`closure')
  errors.
\item As 3C286 is partially resolved to most configurations at all frequencies,
  our observations of this source were used to construct a source model whose
  brightness scale was adjusted to give the correct total flux density given
  in Table~\ref{table:2}.
\item These models of 3C286 were then used by CALIB to determine the antenna
  gains, and to derive the flux density of the calibration source J1150$-$0023 at
  L, C, and X bands.  For the higher frequencies, the flux density of 3C273
  was derived directly, using the longer spacings for which the nuclear core
  is cleanly separable.  
\item For the higher frequency observations, the elevation dependency
  of the antenna gains was determined by ELINT, with the resulting
  dependencies applied to the data from all sources.
\item The time-variable complex gains (amplitude and phase) were then derived,
  using 1150$-$003 for L, C, and X bands, and using 3C273 itself for U, K, and Q
  bands via CALIB, and applied using CLCAL.
\item The polarization calibration program PCAL was used to determine the
  antenna polarization.  
\item The RCP - LCP phase difference was determined by observation of the
  plane of polarized flux for 3C286, and applied to the data using CLCOR.    
\item The data were then split out by source into individual files, applying all
  gains, for subsequent self-calibration and imaging, using AIPS program SPLIT.
\end{enumerate}

Calibration of the 73.8 and 327.5 MHz data followed a similar procedure (minus
the polarization calibration stage), except that at 327.5 MHz, the phase
calibration was based on the global average of observations of 3C286, 3C147,
3C48, and 3C138, while at 73.8 MHz, only Cygnus A was used for calibration.

All subsequent calibration was done on a single-source basis,
separately for each frequency and configuration, utilizing
well-established methodologies of self-calibration.  For sources like
3C273, comprising simple structures with sharp brightness gradients,
these procedures are especially effective.  Although the use of the
spectral line correlator modes reduced the `closure'\footnote{These
  are defined as baseline-based errors which cannot be factored out by
  antenna.} errors to very small values -- typically to 0.1\% and 0.05
degrees -- a baseline-based calibration step, using the AIPS program
BLCAL, was done for some of the frequency/configuration databases, if
the characteristic sign of the effect of these errors was seen in any
map\footnote{For a source located near the equatorial plane, like
  3C273, these errors are manifested as a vertical disturbance up the
  center of the image.}

Self-calibration was done for each frequency and configuration
separately.  Following this, the databases from different
configurations were combined to permit imaging over a very wide range
of spatial scales, and to provide the best dynamic range and
sensitivity at each frequency.  Because the nuclear source of 3C273 is
variable, and the observations for AR334 and AR371 were taken over 2.5
year timescale, the flux density of the nuclear emission (in all three
Stokes' parameters) had to be adjusted to a common value before
combination.  This was done using the AIPS program UVMOD, using the
A-configuration values as the reference.  The flux densities and
polarizations of the nucleus, as a function of time and frequency, are
given in Table~\ref{table:3}.
\begin{table}
\caption{Flux Density and Polarization Variability of 3C273}
\label{table:3}
\centering
\begin{tabular}{c|c|c c c c c c}
\hline\hline
Date&&L&C&X&U&K&Q\\
&&1365&4885&8415&14965&22345&43340\\
\hline
        &I     &32.7&36.0&33.0&28.4&23.8&20.0\\
09/07/95&$\chi$&$-$27 &$-$23 &$-$26 &$-$21 &$-$35 &$-$19\\
        &\%    &0.9 &3.0 &3.4 &5.6 &4.3 &4.6\\
\hline
        &I     &32.8&35.6&30.0&24.3&19.8&16.1\\
03/10/95&$\chi$&$-$28 &$-$21 &$-$26 &$-$24 &$-$29 &$-$32\\
        &\%    &1.0 &3.1 &3.0 &5.8 &6.0 &3.4\\
\hline
        &I     &31.7&36.1&26.9&22.3&24.3&30.8\\
01/03/96&$\chi$&$-$34 &$-$23 &$-$26 &$-$30 &$-$40 &$-$40\\
        &\%    &0.8 &3.0 &4.2 &6.1 &4.1 &2.9\\
\hline
        &I     &    &29.6&26.7&31.5&35.6&37.7\\
01/11/97&$\chi$&    &$-$23 &$-$26 &$-$33 &$-$36 &$-$38\\
        &\%    &    &3.1 &4.7 &3.2 &2.8 &2.1\\
\hline
\end{tabular}
\end{table}

\section{Imaging}

We present here images of the structure of 3C273, with resolutions
chosen to highlight the major structural components of the source.

\begin{figure}
\centering
\includegraphics[width=8.5cm]{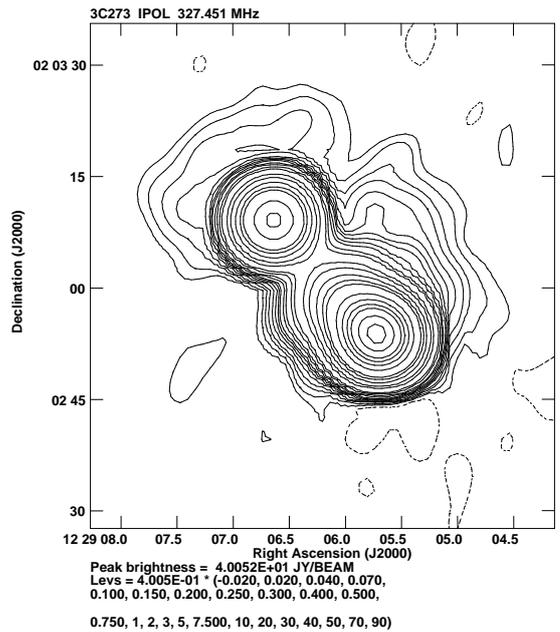}
\caption{3C273 at 327 MHz, with 6 arcsecond resolution.  The extended
  halo emission, slightly offset from the bright emission, is clearly
  visible. }
\label{figure:1}
\end{figure}

\begin{figure}
\centering
\includegraphics[width=8.5cm]{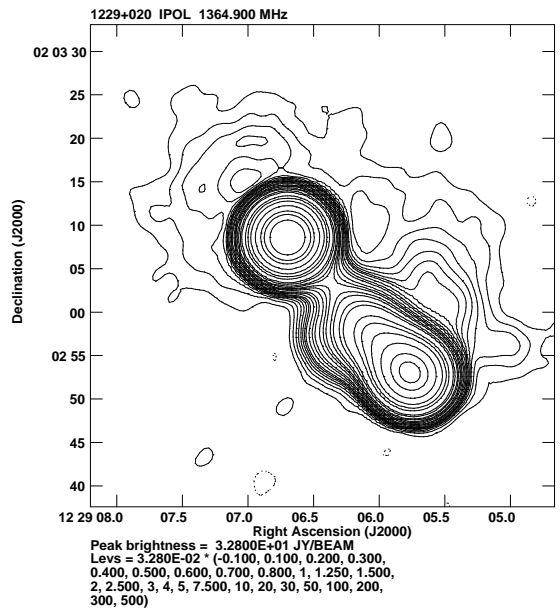}
\caption{3C273 at 1365 MHz, with 4 arcsecond resolution.  The extent of
the halo is similar to that at 327 MHz, although the details of the
emission differ somewhat. }
\label{figure:2}
\end{figure}

\begin{figure}
\centering
\includegraphics[width=8.5cm]{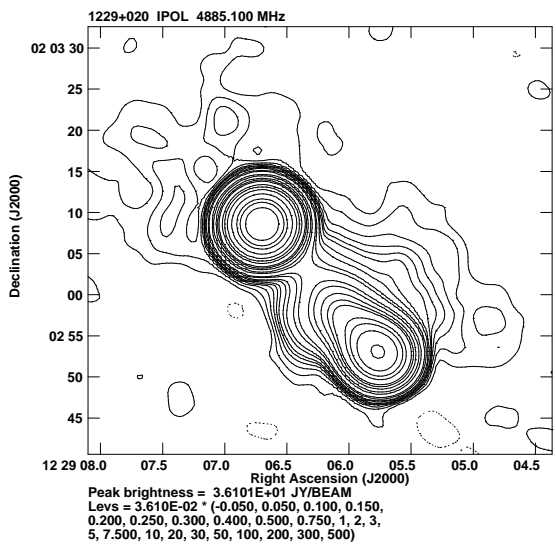}
\caption{3C273 at 4885 MHz, with 4 arcsecond resolution. The overall
  extent of the diffuse emission is similar to that at 327 and 1365
  MHz, but the details differ, most likely due to variations
  introduced by the calibration and imaging process. }
\label{figure:3}
\end{figure}

Low resolution images of the source at $\sim5$ arcseconds resolution
at P, L, and C bands show the presence of a previously unknown diffuse
component of emission.  This emission is of approximate extent 30 by
45 arceconds, centered approximately 10 arcseconds (or 3 kpc) to the
north of the brighter core and jet emission, and aligned with the jet
axis.  Figure~\ref{figure:1} shows the structure at 327 MHz, with 6
arcsecond resolution, figures~\ref{figure:2} and ~\ref{figure:3} show
the structure at 1365 MHz and 4885 MHz with 4 arcseconds resolution.
Imaging at 1465 MHz with 15 arcseconds resolution shows no other
extended emission to our brightness sensitivity level of $\sim$ 0.6
mJy/beam.

The extended emission to the north-west is unpolarized to a level of $\sim$0.1
mJy/beam, implying a maximum fractional polarization of <5\%.  The brighter
extended emission, immediately to the north of the jet shows a marginal
detection of polarized flux, with a maximum fraction of ~$\sim$10\%.

The images made at this resolution clearly show the halo at the lowest
three frequencies (327, 1465, and 4885 MHz).  The differences in the
halo brightness distribution apparent between Figures 1 and 2 should
not be interpreted as evidence for significant spectral variations, as
the process of self-calibration, and in particular, that of
baseline-based calibration, are likely to have generated these
variations.  We do, however, believe the detection of the halo to be
secure.

The three major components -- core, jet, and halo, can be separated by
suitable integration of the image brightness, with the resulting total
flux densities for these components given in Table~\ref{table:4}.  We
emphasize that these values of the halo flux are approximate -- a 30\%
error should be assumed.
\begin{table}
\caption{Core and Extended Flux Densities for 3C273}
\label{table:4}
\centering
\begin{tabular}{c c c c}
\hline\hline
Frequency&Core&Jet&Halo\\
MHz&Jy&Jy&Jy\\
\hline
73.8&9.1&151.5&?\\
327.5&12.7&52.8&0.80\\
1365&32.80&17.2&0.30\\
4885&36.10&5.75&0.15\\
8415&33.12&3.49&0.04\\
\hline
\end{tabular}
\end{table}
This extended halo of emission -- unknown before these observations --
may represent the lobes of a low-luminosity `FR1' radio source.  In
this interpretation, the nuclear core and the one-sided jet represent
Doppler-brightened emission from a relativitic flow aligned close to
the line of sight.  This interpretion is explored by \citet{PK16}.

\begin{figure}
\centering
\includegraphics[width=8.5cm]{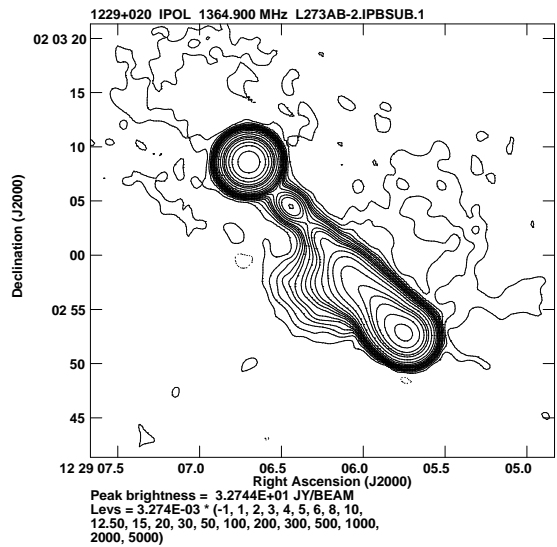}
\caption{3C273 at 1365 MHz, with 2 arcsecond resolution.  The inner and outer
  jets are now visible, while the faint halo is now nearly resolved out.  The
  brighter extension to the SE of the outer jet is clearly visible.  }
\label{figure:4}
\end{figure}

\begin{figure}
\centering
\includegraphics[width=8.5cm]{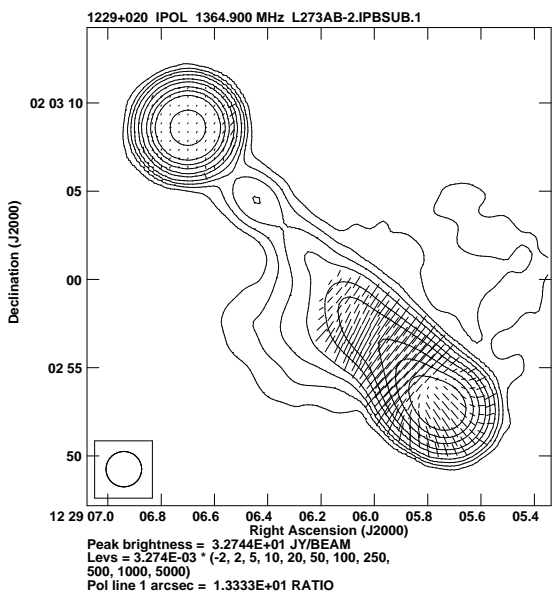}
\caption{The polarization of 3C273 at 1365 MHz with 2 arcsecond resolution.
  The plotted E-vectors show the orientation of the plane of linear
  polarization, their lengths are proportional to the degree of
  polarization. The observed pattern shows that the projected magnetic field
  is along the jet, and wraps around the end.  This pattern, with the B-field
  aligned with the edges of the emission, is the general rule for
  synchrotron-emitting extragalactic radio sources.  The extended emission to
  the SE of the jet is polarized at a level at or less than 15\%.  }
\label{figure:5}
\end{figure}

It will be noted from Figures 1 and 2 that there is a modest `bulge'
of emission on the southern side of the source, about midway down the
jet.  This structure, which was previously seen by imaging with
MERLIN, is more clearly shown at 2 arcsecond resolution, in
Figure~\ref{figure:4}.
In Figure~\ref{figure:5} we show the linear polarization of the
emission at a 2 arcsecond resolution at 1365 MHz.

\begin{figure}
\centering
\includegraphics[width=8.5cm]{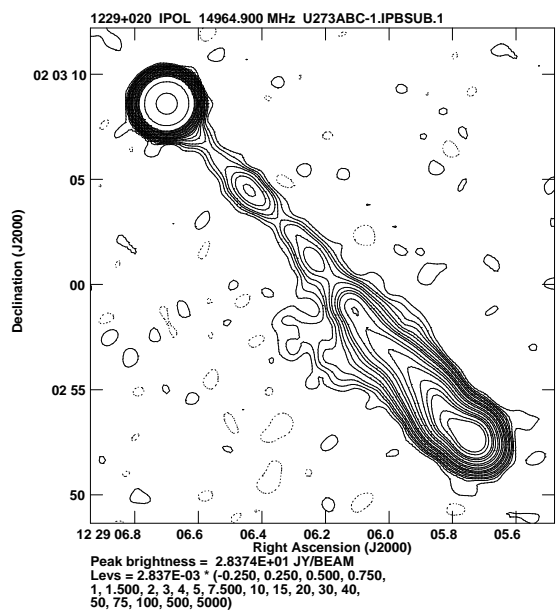}
\caption{3C273 at 14965MHz, with 1 arcsecond resolution. The extended
  emission is nearly resolved out, while details of the inner jet are
  becoming clear. }
\label{figure:6}
\end{figure}

Increasing the resolution by another factor of two resolves out the
extended emission completely, and reveals detailed structures of the
jet emission itself.  Figure~\ref{figure:6} shows the emission with 1
arcsecond resolution at 14965 MHz.  Notable here is the presence of a
bright, straight, apparently unresolved inner jet, extending halfway
between the nuclear core and well-resolved outer jet.  

The inner jet has previously been detected in observations by MERLIN
\citep{D85} and the VLA \citep{C93}, but with poorer
sensitivity that the results shown here.  Optical emission from this
inner jet has been reported from HST observations by \citet{M03}.

\begin{figure}
\centering
\includegraphics[width=8.5cm]{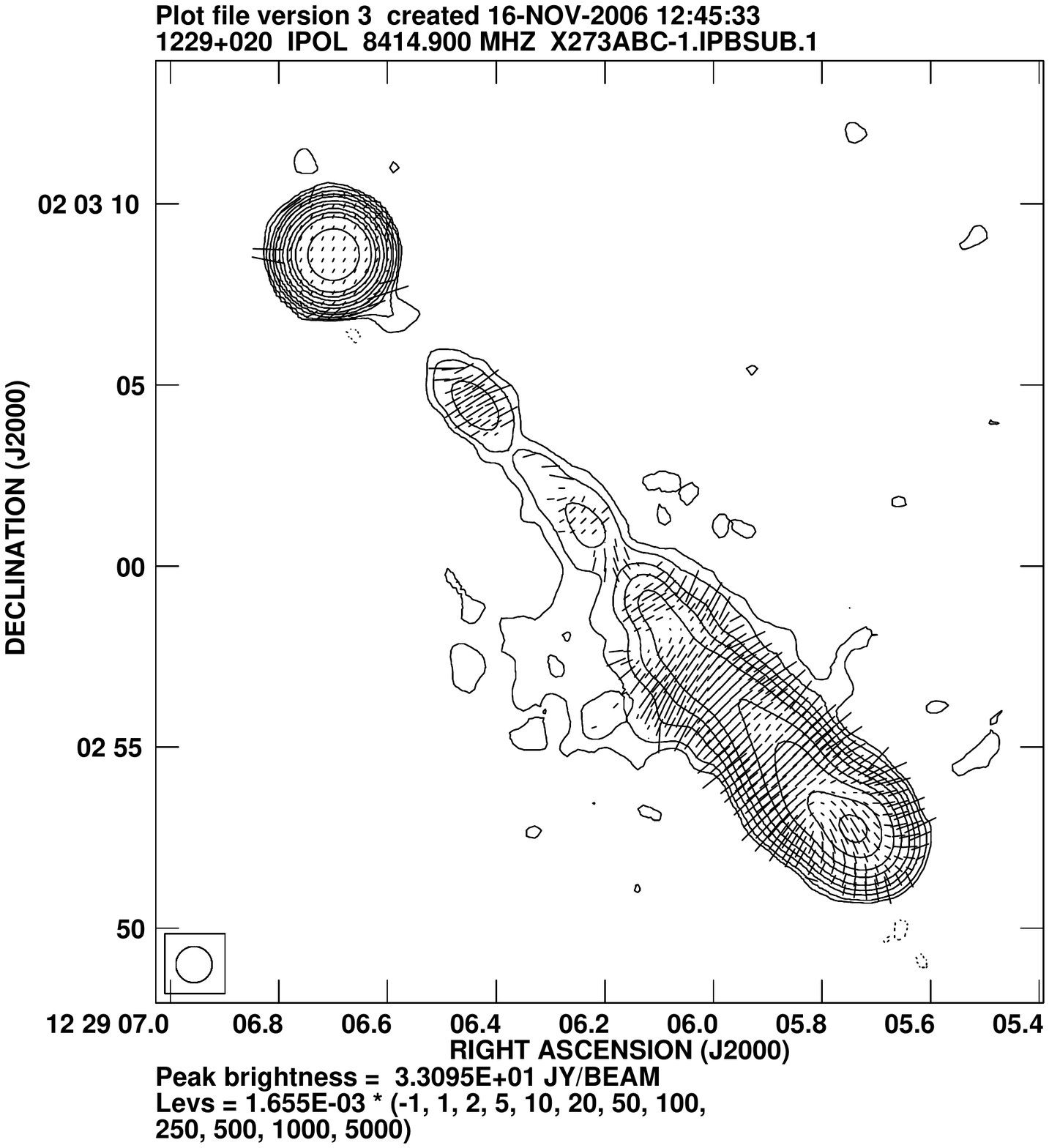}
\caption{The polarization of 3C273 at 8415MHz, with 1 arcsecond
  resolution. The projected magnetic field orientation is orthogonal
  to these plotted E-vectors, and accurately follows the brightness
  isocontours. }
\label{figure:7}
\end{figure}

Figure~\ref{figure:7} shows the polarization at 8415 MHz with 1 arcsecond
resolution.  The inner jet polarization is quite low, typically 5 to 10 \%,
with the B-field direction along the jet.  The outer jet shows considerable
variation in polarization structure, with a high degree of polarization
($\sim$30\%) along the edges, and much lower values throughout the center.  

\begin{figure}
\centering
\includegraphics[width=7cm]{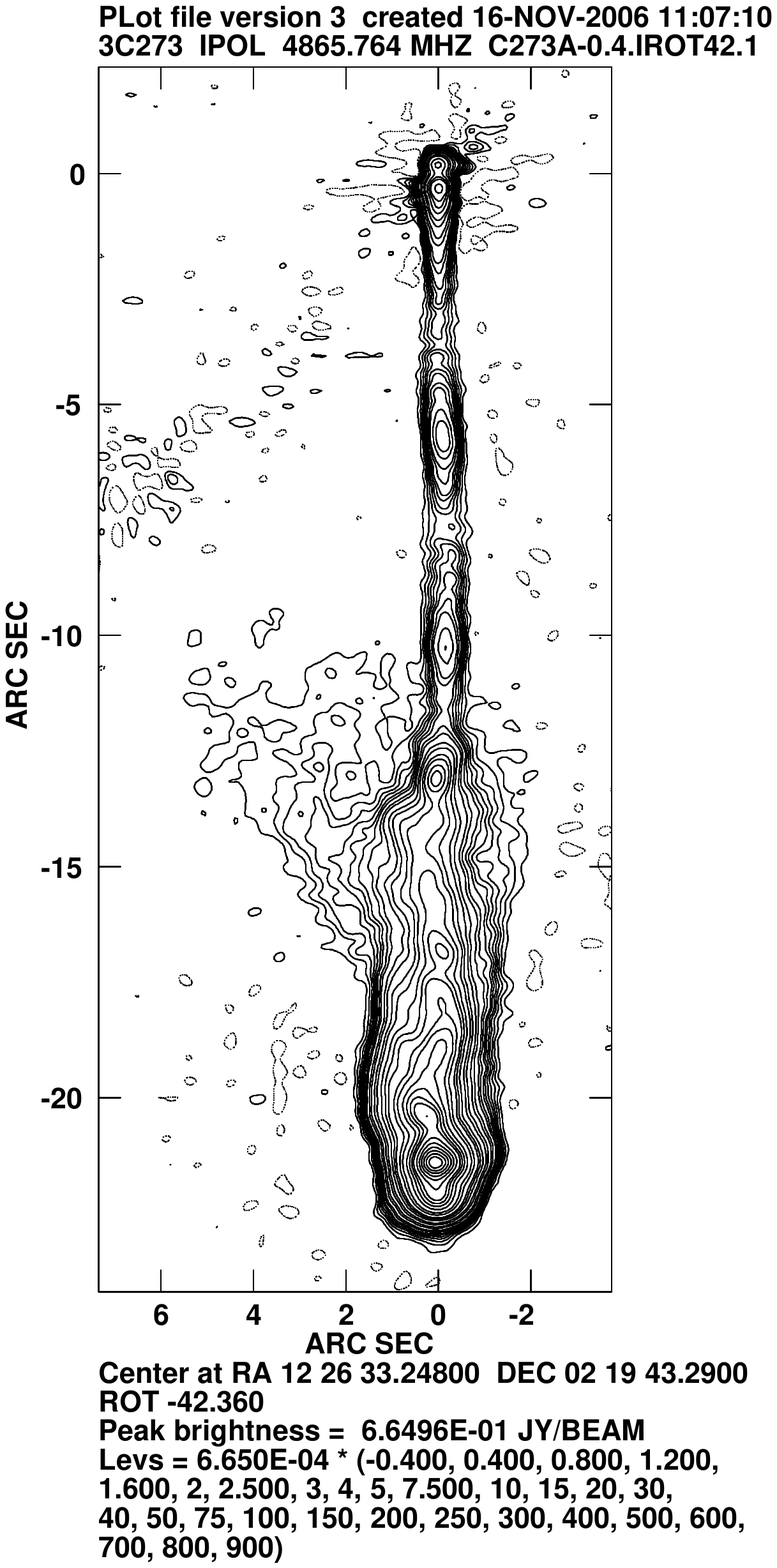}
\caption{3C273 at 4865 MHz and 0.40 arcsecond resolution, rotated CW by 42.36
  degrees. The unresolved nuclear component has been removed, leaving a 30 mJy
  residual at the top to  mark the nuclear position.  Notable here is
  the high brightness of the inner jet, which is transversely unresolved.}
\label{figure:8}
\end{figure}

The structures of the inner and outer jets are strikingly different, as
illustrated in a very deep image, taken at 4866 MHz, with 0.4 arcsecond
resolution, shown in Figure~\ref{figure:8}.    
Some important features of the jet structure seen in the figure are
noted here:
\begin{itemize}
\item The inner jet is nearly straight, and is transversly unresolved for the first
  13 arcseconds of its length.
\item There are three elongated enhancements in brighness along this inner
  portion, separated by 5 arcseconds.  
\item These three elongated enhancements are not perfectly colinear -- the outer
  two are tilted, in opposite directions, by about 5 degrees.  The innermost
  enhancement is perfectly colinear with the brightest hotspot near the end of
  the outer jet.  
\item An emerging, unresolved, knot of emission is seen 550 milliarcseconds
  outside the nuclear position.  This structure is to be identified
  with the innermost jets seen on milliarcsecond scales by the VLBA by \citet{ZT05}   
\item The inner jet abruptly transitions to a wider ($\sim$3 arcseconds) outer
  jet, 13 arcseconds from the nucleus.  
\item The outer jet exhibits considerable variations in structure, including
  an oscillatory center brightness with 6 arcseconds period, and
  $\sim$1arcecond amplitude.  
\item There is a prominent knot (`H2') of emission oriented transverse to the
  jet axis, 22 arcseconds from the nucleus.  Emission is detected for
  two more arcseconds beyond this knot ('H1').  
\end{itemize}

\begin{figure}
\centering
\includegraphics[width=8.5cm]{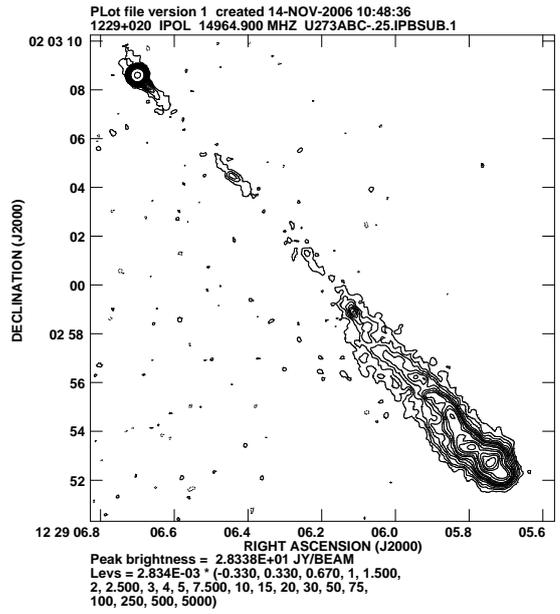}
\caption{3C273 at 14965 MHz, with 0.25 arcsecond resolution. The
  oscillatory structure of the outer jet is now more clearly evident. }
\label{figure:9}
\end{figure}

The oscillatory behavior of the central spine of emission within the
outer jet is more clearly seen with 0.25 arcsecond resolution at 14.9
GHz, as shown in Figure~\ref{figure:9}

\begin{figure}
\centering
\includegraphics[width=8.5cm]{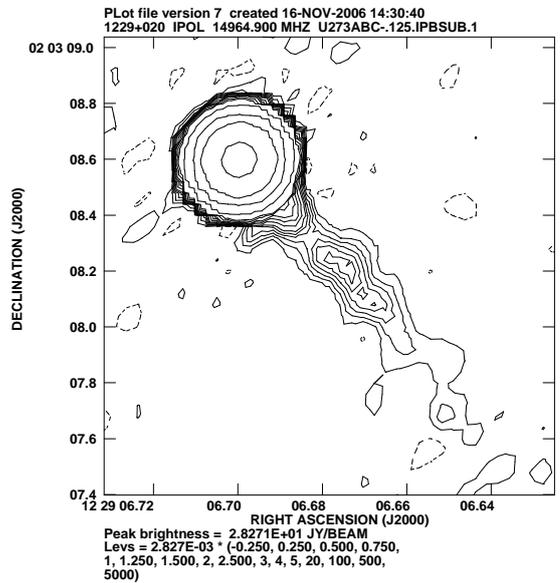}
\caption{The innermost jet, as seen at 14965 MHz, with 125 milliarcsecond
  resolution. The `bulge' in emission immediately to the SW of the
  nucleus is the end of the milliarcsecond-scale jet.}
\label{figure:10}
\end{figure}

\begin{figure}
\centering
\includegraphics[width=8.5cm]{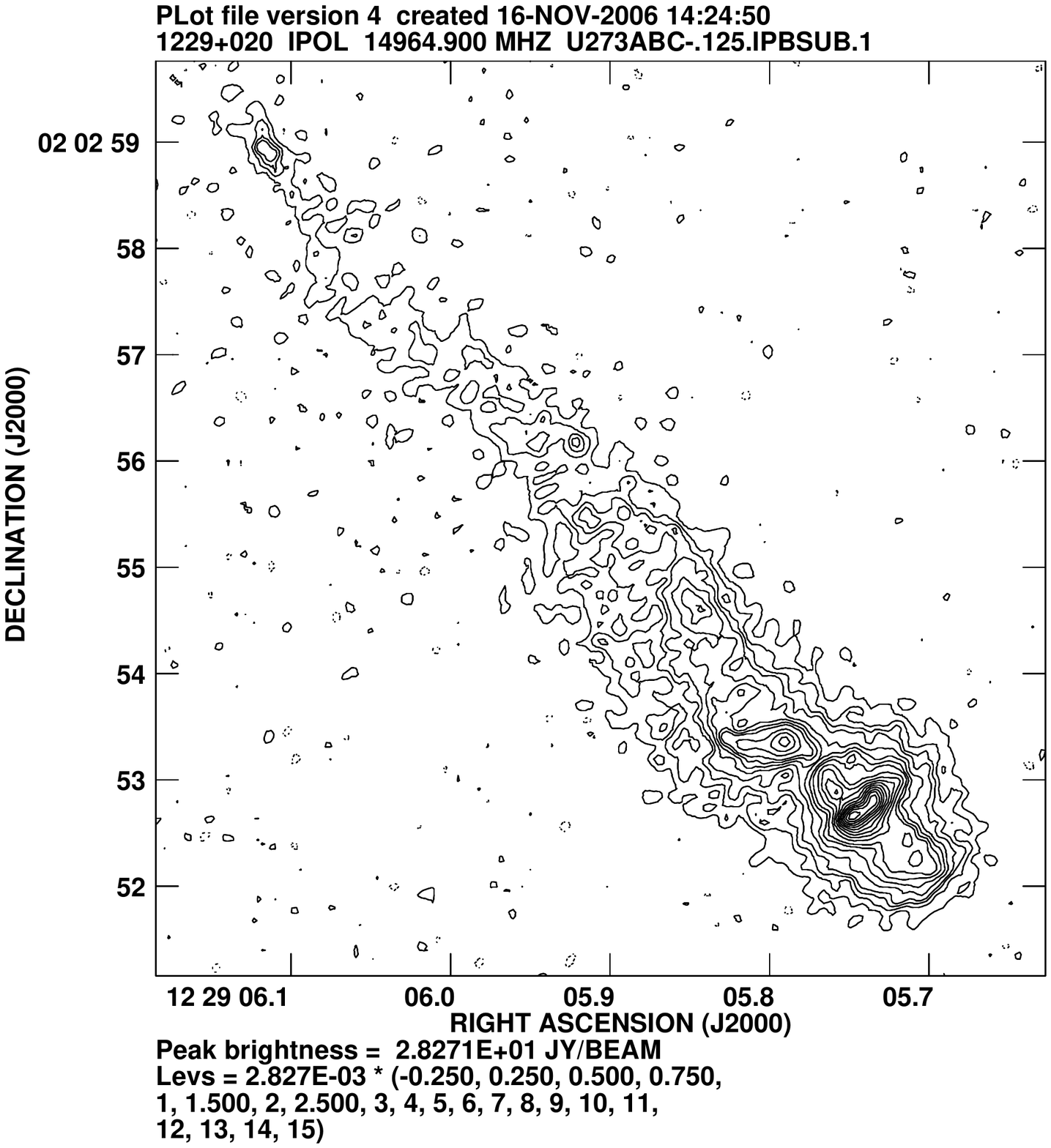}
\caption{The brightest portions of the outer jet, showing the
  brightest hotspot, and the oscillatory ridge leading to it.}
\label{figure:11}
\end{figure}

At 125 milliarcsecond resolution, more details of the brighter components of
the jet become visible.  The innermost jet is clearly visible, as shown in
Figure~\ref{figure:10}.
No polarized flux from this inner jet was detected, as the noise level in this
region is near 5 mJy/beam -- comparable to the jet brightness.  
The outer jet, at this same resolution, is seen in Figure~\ref{figure:11}

\begin{figure}
\centering
\includegraphics[width=8.5cm]{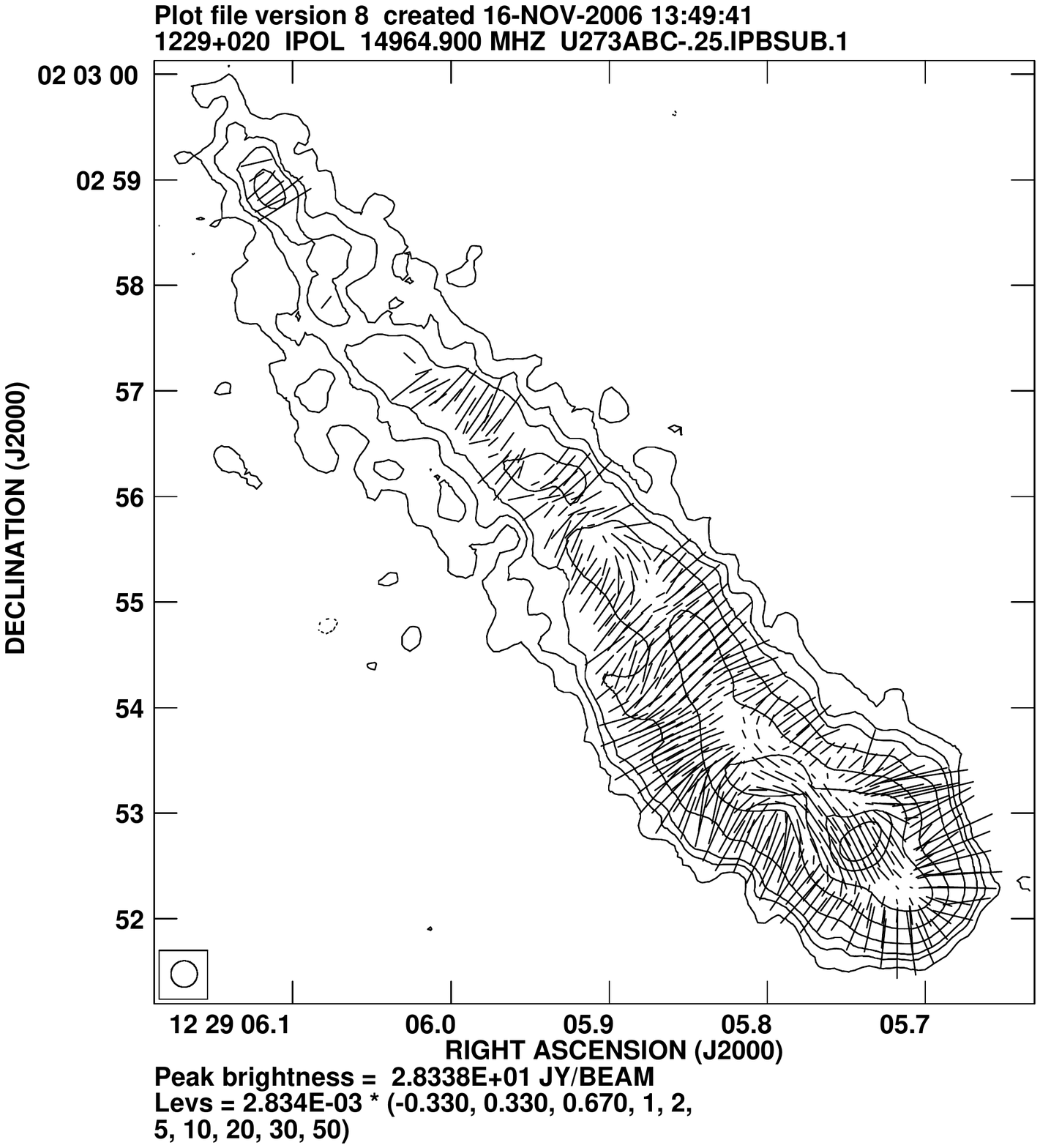}
\caption{The polarization structure of the outer jet at 14965 MHz, with 0.25
  arcsecond resolution.  The length of the vectors indicates the
  fractional linear polarization.  Their orientation is perpendicular
  to the projected jet magnetic field.}
\label{figure:12}
\end{figure}

\begin{figure}
\centering
\includegraphics[width=8.5cm]{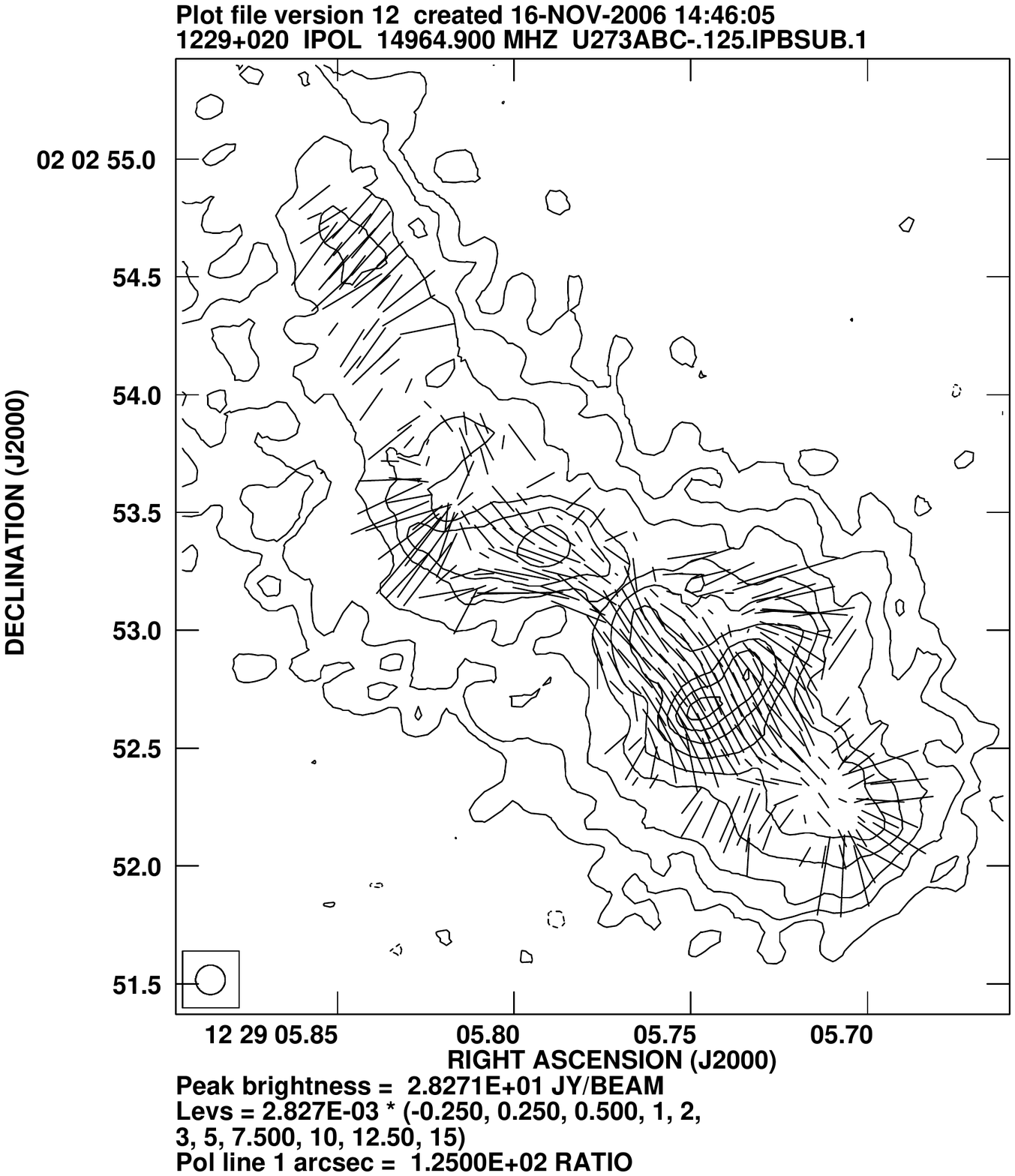}
\caption{The polarization structure of the outer jet at 14965 MHz, with 0.125
  arcsecond resolution. The polarized emission from the more extended
  regions of the jet is not visible due to insufficient sensitivity. }
\label{figure:13}
\end{figure}

The detailed polarization image at 0.25 arcsecond resolution shows a
considerably more complex structure than in total intensity, as shown
in Figure~\ref{figure:12}.  The polarization fraction is generally low
in the interior regions of the jet, and reaches high values -- as high
as 55\% -- along the boundaries.  The projected magnetic field
accurately follows the total intensity boundaries of the jet.
At 0.125 arcseconds resolution, the only detectable polarization
emission at 14.9 GHz is from the brightest structures associated with
the hotspot, as shown in Figure~\ref{figure:13}
The degree of polarizaton in general remains relatively low in these
bright areas -- typically 10\%, but reaches $\sim$40\% at the bright
hotspot, with a magnetic field orientation perpendicular to the jet
main axis.

\begin{figure}
\centering
\includegraphics[width=8.5cm]{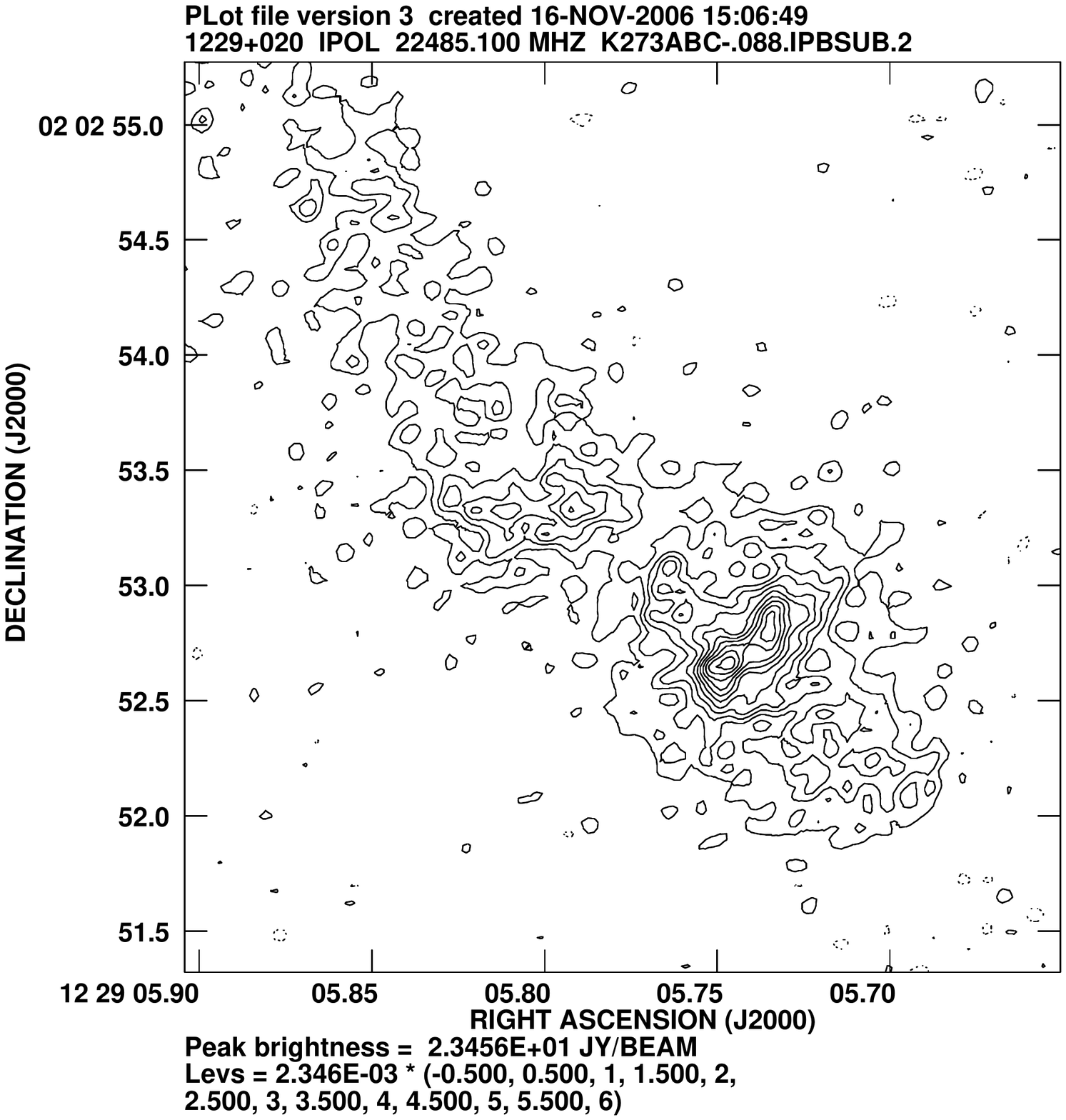}
\caption{The structure of the brightest hotspot at 22345MHz, with
  0.088 arcsecond resolution.  The hotspot is now almost completely
  resolved out, indicating there are no physical structures less than
  $\sim$300pc in size.}
\label{figure:14}
\end{figure}

Concluding this section on the structure of the jet, we show our highest
resolution image -- 88 milliarcseconds at 22345 MHz -- in
Figure~\ref{figure:14}.  The bright hotspot is now fully resolved.
No polarized emission is detected, due to insufficient surface
brightness sensitivity.  

\section{Spectral Index}

\begin{figure}
\centering
\includegraphics[width=8.5cm]{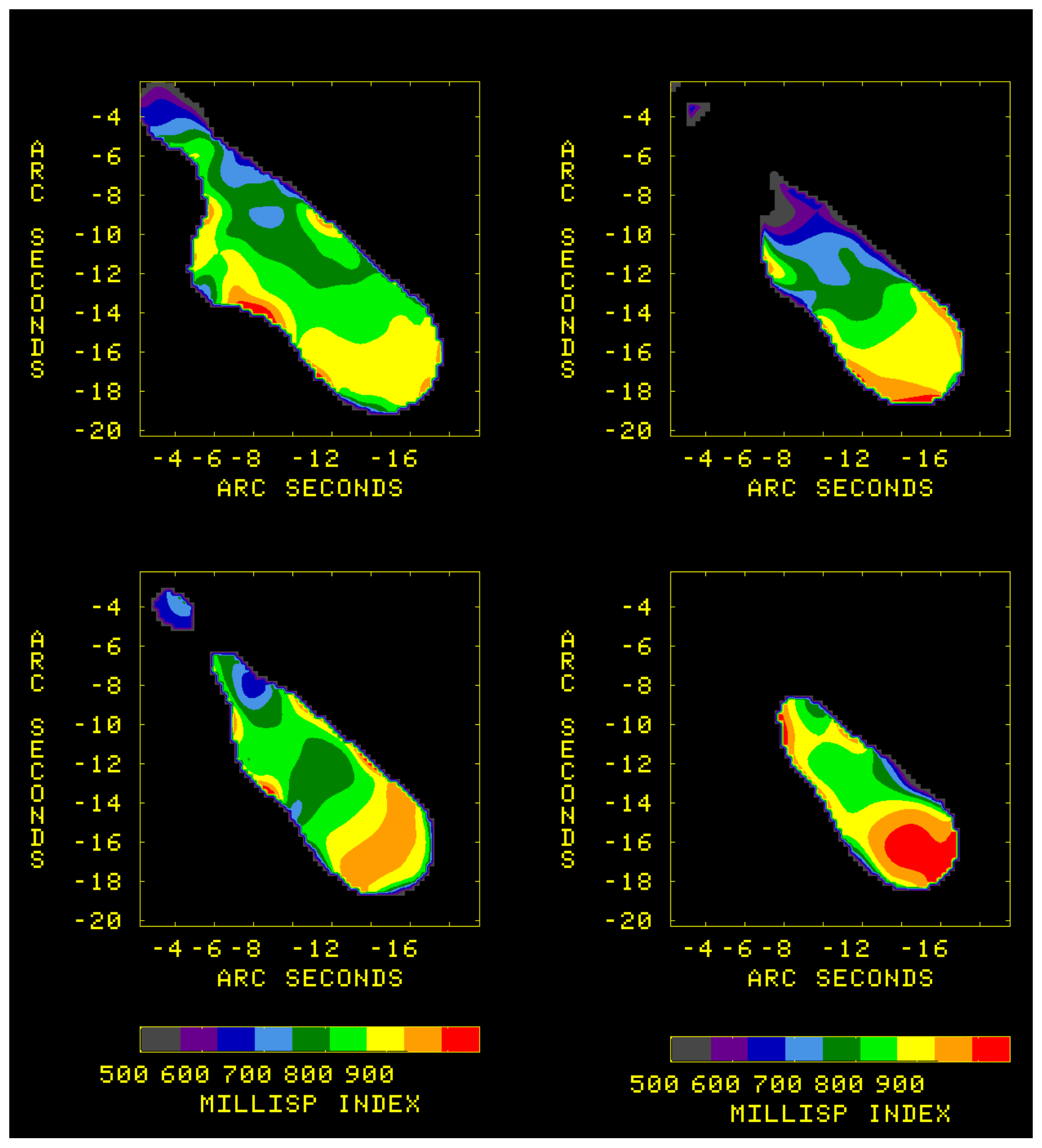}
\caption{A four-panel figure showing the change in spectral index
through the centimeter radio band, at 2.0 arcseconds resolution.
Upper left: $\alpha_{1365}^{4885}$, Upper Right:
$\alpha_{4885}^{8415}$, Lower Left: $\alpha_{8415}^{14965}$, Lower
Right: $\alpha_{14965}^{22345}$.  These show the dramatic steepening
of the spectrum in the region of the bright radio hot spot `H2'.  }
\label{figure:15}
\end{figure}

We show, in Figure~\ref{figure:15} a four-panel spectral index
image\footnote{Defined as $\alpha =
  (\log(B_1)-\log(B_2))/(\log(\nu_2)-\log(\nu_1))$.} at 2.0 arcsecond
resolution, spanning 1.365 through 22.3 GHz.

Figure~\ref{figure:15} shows that the inner jet has a spectral index
near 0.65, and that the outer jet has a uniform spectral index near
0.85.  There is a notable steepening of the radio spectral index in
the vicinity of the brightest hotspot, to -1.0, between 14965 and
22485 MHz. The spectral index in the hotspot vicinity at higher
frequencies continues to steepen, reaching between 22.3 approximately
$\alpha_{23345}^{43340}\sim 1.4$ between 23.3 and 43.3 GHz.  

\begin{figure}
\centering
\includegraphics[height=8.5cm, angle=-90]{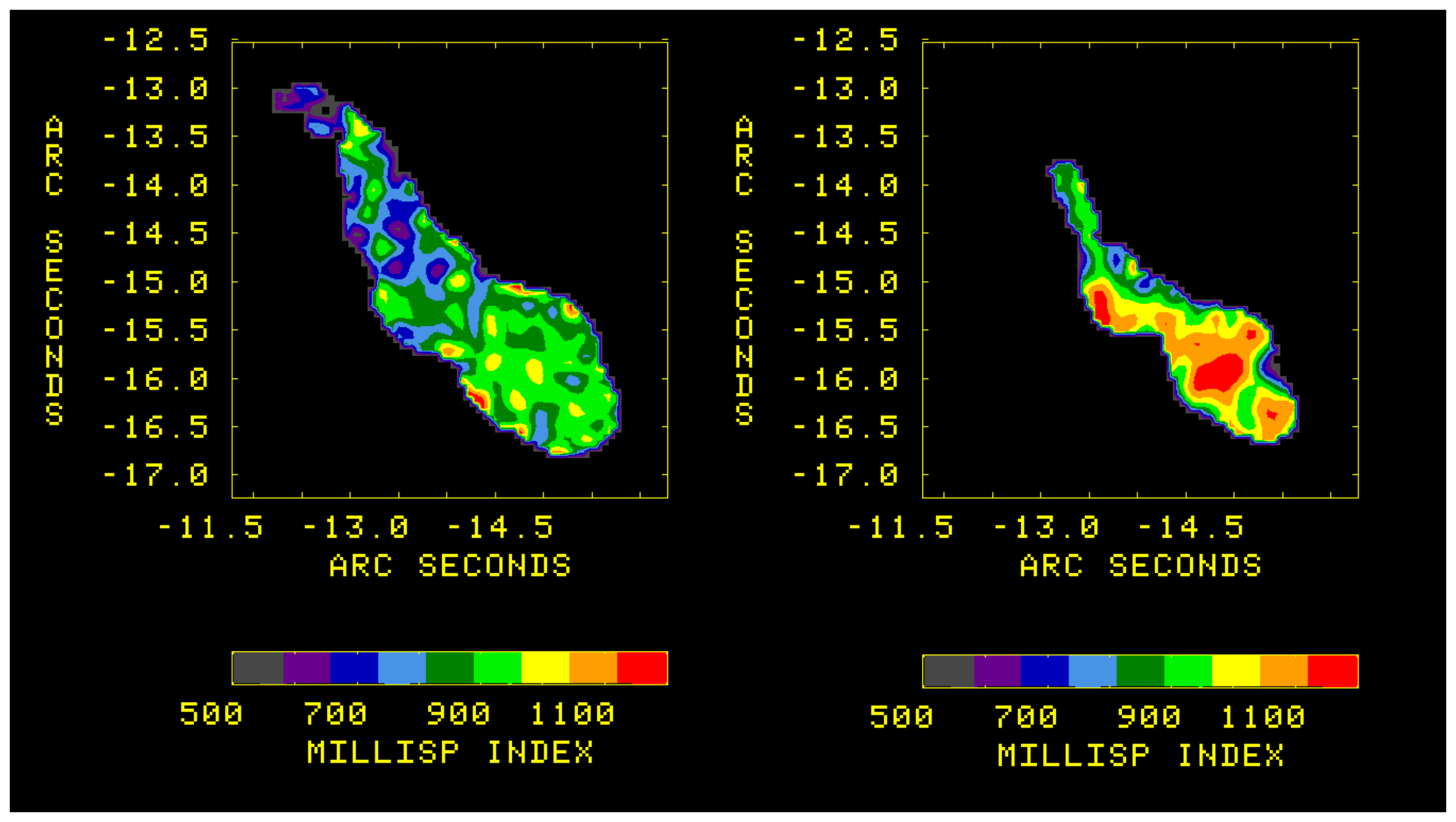}
\caption{A two-panel figure showing the change in spectral index
  through the centimeter radio band at 0.25 arcseconds
  resolution. Left: $\alpha_{8415}^{14965}$, Right:
  $\alpha_{14965}^{22345}$.  These show the dramatic steepening of the
  spectrum in the region of the brightest radio hot spot `H2' is not
  confined to the hotspot alone, but extends well beyond, in both
  directions.  }
\label{figure:16}
\end{figure}

The steeping of the spectrum is not a property of the brightest hotspot only,
however, as is made clear in Figure~\ref{figure:16}, showing the
spectral index images between X and K bands at 0.25 arcseconds resolution.  

The spectral index of the low-brightness extended emission to the
north and east could not be determined, as the intensity is too low to
permit an accurate image.  However, the integrated fluxes shown in
Table~\ref{table:4} can be used to provide a rough measure of $\alpha
\sim 0.6$.

The interpretation of these spectral trends is beyond the scope of
this paper.  We refer the reader to the thorough discussion given in
\citet{Jester05}, which incorporates observations extending from the
radio through the ultraviolet.  

\begin{figure}
\centering
\includegraphics[width=8.5cm]{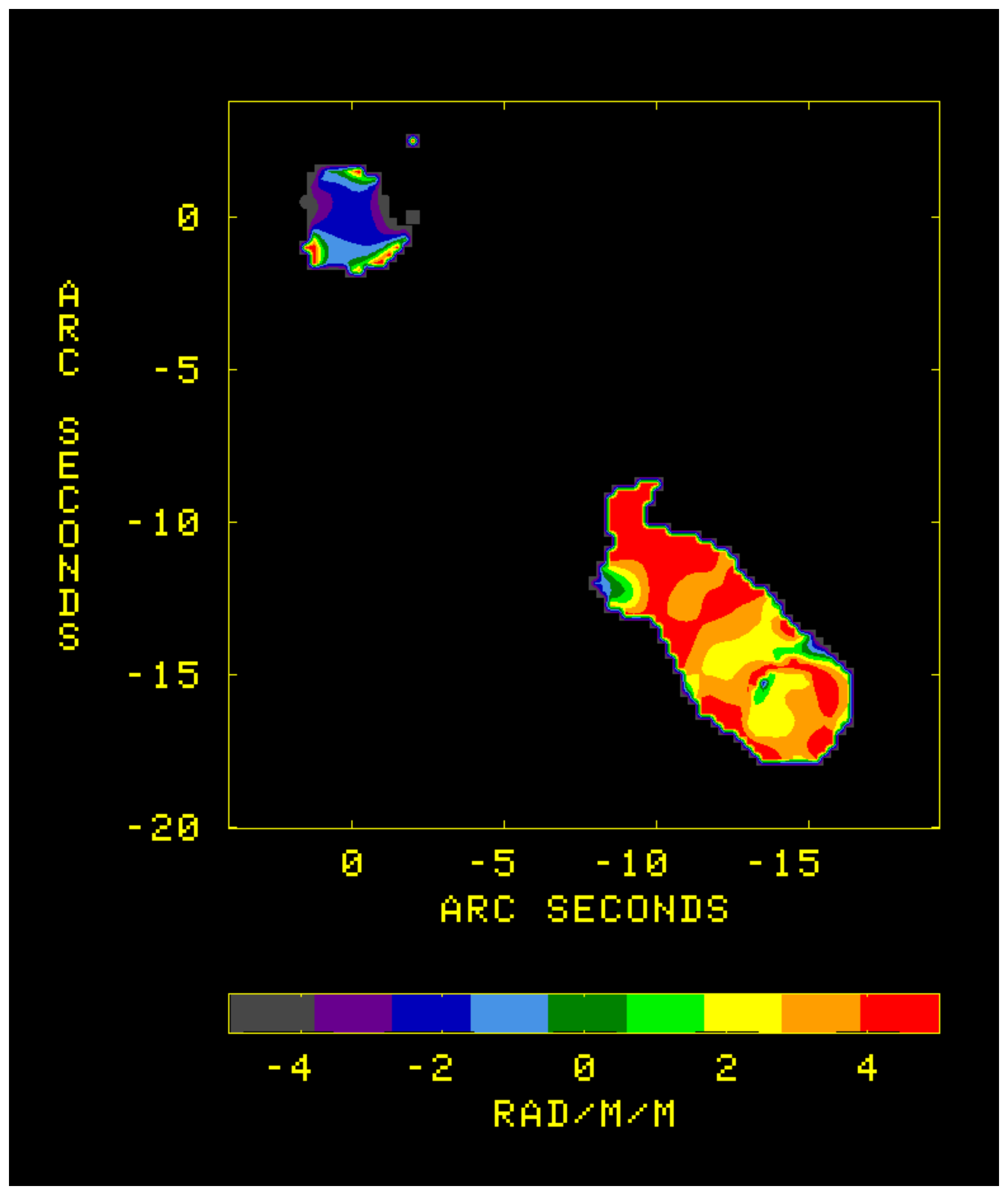}
\caption{The rotation measure as determined from the 1365 and 4885 MHz
  position angle images at 1.4 arcsseconds resolution.  The
  small-scale variations seen in both the jet and nucleus are
  artefacts from the noise.  The jet has a uniform RM both across and
  along its axis. }
\label{figure:17}
\end{figure}

Finally, we note that the polarization images allow a determination of
the rotation measure.  The most sensitive measure is between the
lowest pair of frequencies, 1365 and 4885 MHz.  Figure~\ref{figure:17}
shows the RM at 1.4 arcseconds resolution.  The RM of the entire outer
jet is uniform, with mean value 3.7 $\mathrm{rad}/\mathrm{m}^2$, with
a dispersion of only 1.7 $\mathrm{rad}/\mathrm{m}^2$.  Three is no
discernible gradient either across, or along the jet.  The RM of the
nucleus is slightly less:  $-2.0\:\mathrm{rad}/\mathrm{m}^2$.

\section{Image Archive}
The results shown here are a small fraction of the approximately 500
images available from this project.  We have placed the full suite of
images, in FITS format, in a publically accessible ftp site.  To
access by anonymous ftp, go to `ftp.aoc.nrao.edu', and change
directory to `pub/staff/rperley/3C273'.  The images are named in a
manner which provides sufficient information to identify the
content. In general, files are named thusly: \\[0.3cm]
{\bf bb273cccc-rr.type}\\[0.3cm]
where
\begin{itemize}
\item `bb' are the band(s) utilized for the image
\item `cccc' are the configurations utilized
\item `rr' is the resolution of the restoring beam, in arcseconds.  
\item `type' is the type of image.
\end{itemize}
There are many possibilities for `type'.  These include:
\begin{itemize}
\item A leading `I', `Q' or `U' denotes the polarization.  This is
  normally followed by `CLN', denoting a deconvolved image.  If
  followed by `SUB', the image is a subsection of the original
  (deconvolved) image. If followed by `PB', the image has been
  corrected for primary beam attenuation.  
\item `SPX' denotes spectral index.
\item `DGP' denotes degree of polarization.  
\item `POLA' and `PANG' denote polarization angle.
\item `POLC' denote polarization fraction, corrected for Ricean bias.
\item  Numbers following these last four items give the permitted
  error threshold applied for blanking the output pixels.  
\end{itemize}
All files end in `.1' -- this is a sequence number and adds no
additional information.  

For example, `U273ABC-1.4.UCLN' is a Stokes U clean image
made with U band data, using the A, B, and C configurations, with 1.4
arcseconds resolution. `UX273-1.4.SPX.1.1' in spectral index image
between U and X bands, with 1.4 arcseconds resolution, with an error
threshold of 0.1 in spectral index.  (All pixels with rms error greater
than 0.1 are blanked).  `U273ABC-1.4.DGPSN3.1' is a fractional
polarization image, at U-band, with 1.4 arcseconds resolution, blanked
at a SNR of less than 3.  

\section{Conclusions}
The Very Large Array has been utilized to generate high-fidelity radio
images of the iconic quasar 3C273.  Major features found by this
work include:
\begin{itemize}
\item A previously unknown low-brightness steep-spectrum diffuse halo,
  of extent 30 x 45 arcseconds, primarily on the northern and western
  side of the bright nucleus and jet.
\item A bright, narrow, nearly straight inner jet, of 11 arcseconds
  length, joining the bright nucleus to the outer jet.  The transition
  in width from the narrow to wider jet is abrupt.  
\item The inner jet comprising three elongated regions of enhanced
  brightness, the outer two of which are oppositely tilted by about 5
  degrees from the jet axis, suggestive of an oscillatory or
  sinusoidal underlying structure of period $\sim$ 5 arcseconds.  
\item The outer jet, of width $\sim 2.5$ arcseconds, which also
  comprises an oscillatory structure, with a similar period, but
  significantly larger amplitude.
\item The outer jet is highly linearly polarized, with the
  polarization fraction reaching 55\% along the jet boundaries.  The
  fractional polarization in the central regions of the outer jet is
  much lower.  The projected magnetic field accurately follows the
  lines of constant brightness, including curling around the leading
  edge of the jet.
\item The inner jet is also highly polarized, with the magnetic field
  lines oriented along the jet axis.  
\item The bright radio hotspot `H2' is fully resolved with $\sim$0.09 arcsecond
  (300 pc) resolution.  
\item The spectrum of the outermost regions of the radio jet sharply
  steepens above 5 GHz.  The inner jet's spectrum is flatter than that
  of the outer jet.
\item The rotation measure of the jet is uniform, with no visible
  gradient either across or along the jet.  
\end{itemize}

Despite the detailed information on the structure of this source
provided by this work, much remains uncertain, or poorly determined.
We note, in particular:
\begin{itemize}
\item The structure of the newly-discovered halo is very uncertain.
  While we are confident in the existence of this emission, the
  differences in structures suggested by comparison of Figures 1 to 3
  are unlikely to be real.  
\item The current images are noise limited at all the higher
  frequencies, particularly in polarization.  Hence, details of the
  jet structure and polarization at high frequencies and at high
  resolution are poorly determined.
\item The polarization of the inner jet is only approximately known,
  since the high-fidelity data taken in 1987 and 1991 were in a
  single-polarization mode.  
\end{itemize}
The recent upgrade of the Very Large Array can address all these, and
other issues.  With the dramatic increase in bandwidth, improved
receiver sensitivies, and especially with the implementation of a much
more powerful correlator, much superior imaging of this source -- and
other such sources -- is easily in range.
 
\begin{acknowledgements}
  RAP is pleased to acknowledge the generosity and hospitality of the
  Max-Planck Institut fur Astronomie, and in particular that of
  Hermann-Josef R\"oser, during his many visits to Heidelberg, where
  most of this work was done.  RAP also thanks Jean Eilek and Brian
  Punsly for helpful comments on the text.  
\end{acknowledgements}


\begin{thebibliography}{}
  
\bibitem[Baars et al. (1966)]{Baars} Baars, J.W.M., Genzel, R., Pauliny-Toth, I.I.K., \&
  Witzel, A. 1977, A\&A, 61, 99
\bibitem[Conway et al. (1993)]{C93} Conway, R.G., Garrington, S.T.,
  Perley, R.A., and Biretta, J.A. 1993, A\&A, 267, 347
\bibitem[Davis et al. (1985)]{D85} Davis, R.J., Muxlow, T.W.B., \&
  Conway, R.G. 1985, Nature, 318, 343
\bibitem[Jester et al. (2005)]{Jester05} Jester, S., R\"oser, H.-J., Meisenheimer, K., and
    Perley, R. 2005, A\&A, 431, 477
\bibitem[Martel et al. (2003)]{M03} Martel, A.R., Ford, H.C., Tran,
  H.D., Illingworth, G.D., Krist, J.E., White, R.L., Sparks, W.B.,
  Gronwall, C., Cross, N.J.G., Hartig, G.F., Clampin, M., Ardila,
  D.R., Bartko, F., Ben\'itez, N., Blakeslee, J.P., Bouwens, R.J.,
  Broadhurst, T.J., Brown, R.A., Burrows, C.J., Cheng, E.S., Feldman,
  P.D., Franx, M., Golimowski, D.A., Infante, L., Kimble, R.A.,
  Lesser, M.P., McCann, W.J., Menanteau, F., Meurer, G.R., Miley,
  G., Postman, M., Rosati, P., Sirianni, M., Tsventanov, Z.I., and
  Zheng, W. 2003, A.J., 125, 2964 
\bibitem[Perley \& Butler (2013)]{PB13} Perley, R.A., and Butler, B.J. 2013, ApJS, 19
\bibitem[Punsly \& Kharb (2016)]{PK16} Punsley, B., and Kharb, P.  submitted to ApJL.
\bibitem[Schmidt(1963)]{S63} Schmidt, M. 1963, Nature 197, 1040
\bibitem[Zavala \& Taylor (2005)]{ZT05} Zavala, R.T., and Taylor,
  G.B. 2005 ApJ, 626, L73
\end{thebibliography}
\end{document}